\documentclass[aps,twocolumn,showpacs,amsmath,amssymb,apl,reprint]{revtex4-1}

\usepackage{graphicx}%
\usepackage{dcolumn}
\usepackage{bm}
\usepackage{braket}
\def\E{{\mathcal{E}}}
\def\T{{\mathcal{T}}}
\begin{document}

\title{Effect of plasma formation on the double pulse laser excitation of cubic silicon carbide}
\author{T. Otobe$^1$, T. Hayashi$^2$, and M. Nishikino$^1$}
\affiliation{$^1$Kansai Photon Science Institute, National Institutes for Quantum and Radiological Science and Technology (QST), Kyoto 619-0215, Japan\\
$^2$ Graduate School of Engineering, Department of Mechanicak Engineering, Kyushuu University,  Fukuoka, 819-0395, Japan}
\begin{abstract}
We calculate the electron excitation in cubic silicon carbide (3C-SiC) caused by the intense femtosecond laser double pulses 
using time-dependent density functional theory (TDDFT).
We assume the electron distributions in the valence band (VB) and the conduction band (CB) based on three different approaches to determine the 
dependence of the plasma  that is formed on the excitation by the first pulse. 
First, we consider the simple double pulse irradiation, which does not include the electron-electron collisions and relaxation.
Second, we consider the partially thermalized electronic state, in which the electron temperatures and numbers in the VB and the CB are defined independently.
This assumption corresponds to the plasma before the electron-hole collisions becomes dominant.  
The third approach uses the fully thermalized electron distribution, which corresponds to a timescale of hundreds fs.
Our results indicate that the simple double pulse approach is the worst of the three, 
and show that the plasma formation changes the efficiency of the excitation by the second pulse.
When the electron temperature decreases, the laser excitation efficiency increases as a result.
\end{abstract}
\maketitle
Processing of solid materials using  femtosecond laser pulses
has attracted considerable interest for potential application
to high-precision processing technology. 
\cite{Chichkov,Stuart,Liu,Lenzner,Geissler,Lenzner00,Sudrie,Doumy,Amoruso,Gattass08, Reif10, Gamaly11,Chimier} 
Because a femtosecond
laser pulse can deposits large amounts of energy into solid
materials within a much shorter time than conventional spatial diffusion of
thermal energy to the exterior of the irradiated spot, we can
process materials with small thermal denaturation outside of the
irradiated volume.\cite{Gattass08, Reif10, Gamaly11}

The peak intensity of a femtosecond laser  pulse can be very high
when compared with that of longer laser pulses and is thus suitable for processing of wide gap materials and dielectrics by nonlinear processes.
However, the high peak laser intensity laser sometimes induces the damage  in optical elements, such as lenses, gratings, and mirrors.
Therefore, the laser processing efficiency at the peak laser intensity is important.

Recently, double-pulse irradiation has been proposed as a new approach for efficient laser processing \cite{Shi14, Sandra15, Sandra15-2,Furukawa16}.
We found that the double- or multi-pulse irradiation reduces the ablation threshold intensity in SiC \cite{Hayashi16,Hayashi16-2}.
Because the timescales of the fast electron-electron and hole-hole relaxation are of the order of 100 fs, 
and the carrier relaxation time is of the order of hundreds fs \cite{Michael17}, features of the formed plasma 
are also important in multi-pulse processing.

In this work, we simulate the electron excitation processes in the 3C-SiC caused by the double femtosecond laser 
pulses using the time-dependent density functional theory (TDDFT) with a real-time real-space approach \cite{Runge84, Bertsch00,Otobe08}.   
SiC is one of the important material as a foundation base in the next generation, because of its wide bandgap ($2\sim3$ eV),  temperature resistance, good thermal conductivity as well as impact resistance \cite{Morkoc94, Matsunami98}.
However, for its hardness and chemical, mechanical stability, SiC is difficult to be processed.
Therefore, efficient processing of SiC by femtosecond laser pulses will be an important technique \cite{Choi16}.
 
To explain the effect of the relaxation process, we assume that the  thermalized electron and hole states in the conduction and the valence bands 
 can be expressed by using some Fermi distribution functions. 
 We assume the one temperature for all electronic states \cite{Sato15} and  individual two temperatures for 
 the conduction and the valence band.
 The first assumption corresponds to a timescale of  $\sim 1$ ps, and the second assumption  corresponds to a timescale of few tens to hundreds of  fs.

In real-time TDDFT, we describe the electron dynamics in a unit  
cell of a crystalline solid under a spatially uniform electric field $\E(t)$. 
By treating the field using a vector potential, we obtained 
$\vec A(t)=-c\int^t dt' \vec \E(t')$. 
We assumed that the laser was linearly polarized,  and that the polarization direction was parallel to the $C-$axis.
The electron dynamics were described by the time-dependent Kohn-Sham (TDKS) equation \cite{Runge84}, which is a  fundamental equation of TDDFT.   
 Real-time  calculations were performed using the  SALMON (Scalable Ab-inito simulator for Light-Matter interactions in Optics and Nanoscience) 
 program package\cite{SALMON}. 
We modified SALMON to treating the temperatures of the valence and the conduction bands individually.  
In this work, we used the modified Becke-Johnson (mBJ) exchange potential \cite{mBJ,mBJ2} given as Eqs. (2)-(4) in Ref. \cite{mBJ2} 
with a  local-density approximation (LDA) correlation potential \cite{LDA} under an adiabatic approximation.
The $c$-value in the mBJ potential was set at 1.35 to reproduce the optical band gap of 3C-SiC (6eV). 

A cubic unit cell containing four silicon atoms and four carbon atoms was discretized into Cartesian grids containing $16^3$ points.
The $k$ space was also discretized into $ 16^3$ grid points. 
 We used a time step of 0.04 a.u.

The laser electric field $\E(t)$ was assumed to be
\begin{equation}
\E(t)=
\begin{cases}
\E_0 \sin^2\left(\pi \frac{t}{\T_p}\right)\sin(\omega_0 t) & 0<t<\T_p \\
0& \T_p<t< \T_e, \label{eq:field}
\end{cases}
\end{equation} 
where $E_0$ is the maximum electric field amplitude and  $\omega_0$ is the center of  the laser frequency ($\omega_0=1.55$ eV).
 $E_0$ is related to the incident laser field ($\E_{\rm in}$) by $\E_0=2/(1+\sqrt{\varepsilon})\E_{\rm in}$, where $\varepsilon$ is the dielectric function at $\omega_0$ \cite{yabana12}.
The pulse length $\T_p$ was set at six optical cycles (16.2 fs), and the polarization was parallel to the [0,0,1] direction.

We used the LDA functional to calculate the time-dependent total energy, $E_{tot}(t)$, because the mBJ  potential
does not have an energy functional.
The absorbed energy $E_{ex}$ is defined as the difference of the $E_{tot}(t)$ between 
the initial and final values of $E_{tot}(t)$,  where $E_{ex}=E_{tot}(t=\T_e)-E_{tot}(t=0)$.

\begin{figure} 
\includegraphics[width=90mm]{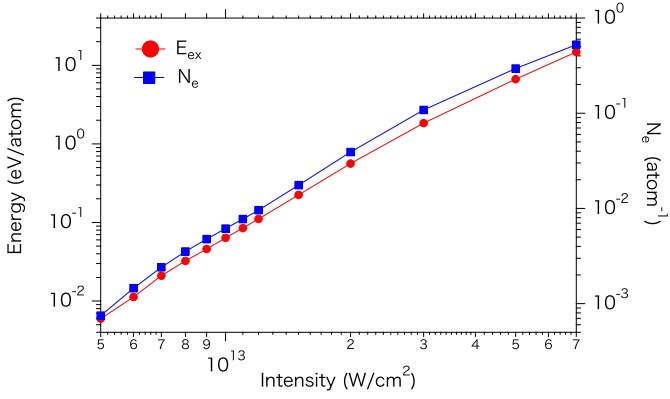} 
\caption{\label{fig2} 
Laser intensity dependences of the energy absorption ($E_{ex}$) and the number of the excited electrons ($N_e$) after the single laser pulse ends. } 
 \end{figure} 
 
The hole density is defined by the projection of the time-dependent wavefunction at $t$ ($u^{\vec{k}}_{i}(t)$) onto the initial state given by 
$N_{\rm e} (t)= \frac{1}{V}\sum_{\vec{k},ii' =occ} \left( \delta_{ii'} -  
\vert \langle \Phi^{\vec{k}+\frac{e}{c}\vec{A}(t)}_{i} \vert u^{\vec{k}}_{i'}(t) \rangle \vert^2 \right)$,
where $i$ and $i'$ are the band indices of the orbitals for the initial and time-dependent states, respectively, 
and $\Phi^{\vec{k}+\frac{e}{c}\vec{A}(t)}_i$ is the wavefunction of the initial state for which the Bloch wave vector is 
shifted by the laser field $\vec{A}(t)$. \cite{Otobe08}

Individual  projections to the initial states given by  
$O^{\vec{k}}_{j}(t) = \frac{1}{V}\sum_{i'=occ }  
\vert \langle \Phi^{\vec{k}+\frac{e}{c}\vec{A}(t)}_{j} \vert u^{\vec{k}}_{i'}(t) \rangle \vert^2$ ,
give the occupation of state $j$. 
We prepare adequate unoccupied states in the conduction band, typically comprising 100 states for each $\vec{k}$, to calculate the overall electron distribution.  
The reduced internal energies in the conduction and valence bands, given by $U_{c}$ and $U_{v}$, respectively, can be defined as 
\begin{equation}
U_{v(c)}=\sum_{\vec{k},i=v(c)} O^{\vec{k}}_i(\T_e)\epsilon^{\vec{k}}_i 
\end{equation}
where $v(c)$ represents states in the valence (conduction) band, and $\epsilon^{\vec{k}}_i$ is the energy eigenvalue of the $i$-th state.

\begin{figure} 
\includegraphics[width=90mm]{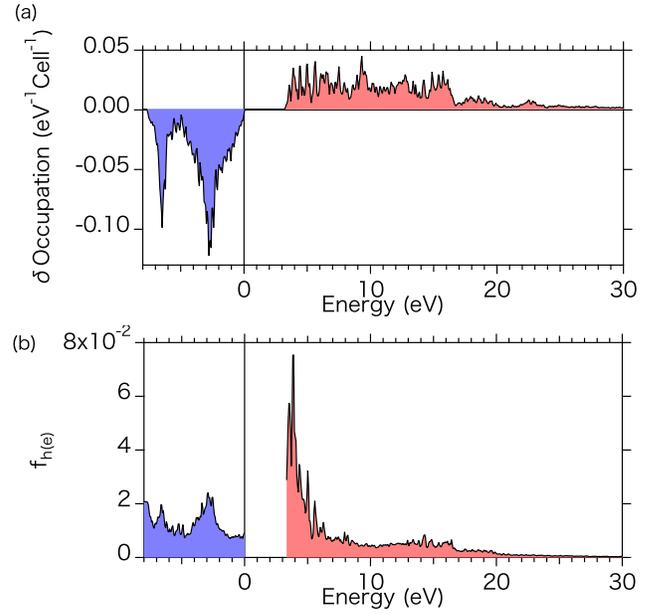} 
\caption{\label{fig3} 
(a) Change in  the electron distribution after the laser pulse ends shown as the DoS. 
The origin for the energy is at the top of the valence band. 
(b) Filling rates of the holes and electrons.} 
\end{figure}

Figure~\ref{fig3} (a) shows the change in the
 electron distribution, $\delta O^{\vec{k}}_i=O^{\vec{k}}_i(\T_e)-O^{\vec{k}}_i(t=0)$, in the form of  the density of states (DoS) 
for the case of a laser intensity of $2 \times 10^{13}$ W/cm$^2$.
A broad distribution in the conduction band and specific peaks in the valence band are shown.
The hole (electron) filling rate, $f_{h(e)}$, is also shown in  Fig.~\ref{fig3} (b).

Three possible assumptions can be used to reproduce the electron distribution after the laser pulse ends. 
First, the collisional process is assumed to be negligible and the electron and hole distributions remain unchanged until the second pulse arrives.
Second, the electrons and holes  are assumed to relax in each of the bands, and the distribution is thermalized.
In general, the electron-electron collision time in the same band is shorter than the corresponding time
between the valence and the conduction bands \cite{Michael17}.
Therefore, this assumption may corresponds to the transient state leading to true thermalization.
Third, the whole system is assumed to be thermalized and the electron distribution can then be expressed using a single temperature,
the corresponds to a much longer time delay than the second assumption.

\begin{figure} 
\includegraphics[width=90mm]{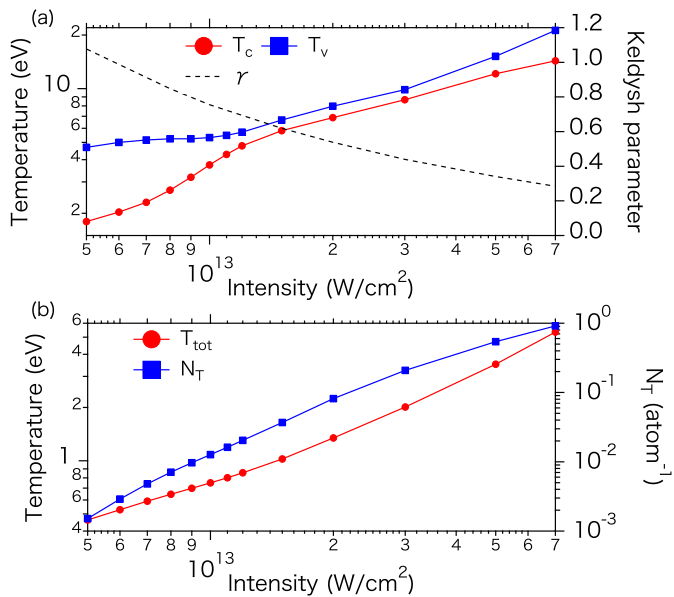} 
\caption{\label{fig4} 
(a) Fitted electron temperature in the valence (blue square) and conduction bands (red circle) as the functions of the laser intensity.
The black dashed line presents the Keldysh parameter, $\gamma$.  
(b) Total temperature ($T_{tot}$) in all bands and electron density in the conduction band ($N_T$).} 
\end{figure}

In the second assumption, we can define the distribution function $F_{v(c)}$  to reproduce
$U_{v(c)}\sim \sum_{\vec{k},i=v(c)}F_{v(c)}(\epsilon^{\vec{k}}_i)\epsilon^{\vec{k}}_i$ at a fixed $N_e$.
In this fitting, both the chemical potential and the electron temperature in each of the bands are treated as parameter.
Figure~\ref{fig4} (a) shows the laser intensity dependence of $T_c$ and $T_v$.
The figure shows a  nonlinearly increase in temperature in the conduction band above $7\times 10^{12}$ W/cm$^2$.
The black dashed line represents the Keldysh parameter ($\gamma$),  which serves as  an index for 
the multiphoton and tunneling excitation processes.
The multiphoton process is dominant when $\gamma \gg 1$, while the tunneling process is dominant when $\gamma \ll 1$.
$\gamma$ acrosses 1 at an intensity of $7\times 10^{12}$ W/cm$^2$.
Therefore the tunneling process increases  the conduction band temperature 
and the temperatures of the valence and conduction bands becomes closer.

Figure~\ref{fig4} (b) shows the fitted electron temperature in all bands ($T_{tot}$)  to reproduce $U_{tot}=U_v+U_c$,
 which corresponds to the third assumption.
In this fitting, we define the single electron temperature. 
Therefore, the number of the excited electrons cannot by fixed artificially. 
The calculated number of excited electrons ($N_T$) is also plotted using  blue squares. 
$N_T$ is approximately twice of $N_e$ at  all intensities.
The figure shows a small kink at an intensity of $1\times 10^{13}$ W/cm$^2$ that corresponds to the tunneling regime, where $\gamma< 1$.

In the next step, we simulated the double pulse excitation.
We assumed that the intensity of the first pulse is $2\times 10^{13}$ W/cm$^2$.
The absorbed energy at an  intensity of  $2\times 10^{13}$ W/cm$^2$ is 0.56 eV/atom, 
which is much lower than the cohesive energy of  6.34 eV/atom \cite{Jiang16,Chang87} 
and higher than the melting point energy of $3.22\times 10^{-2}$ eV/atom \cite{Sace60, Handbook}.
Therefore, ultrafast deformation related to ablation may not occur \cite{Sato15}.  
We confirmed that the $T_c$ is 6.9 eV, and $T_v$ is 8.0 eV in the second assumption.
In the third assumption,  the electron temperature and $N^i_e$ are defined as $T_{tot}=1.35$ eV and $N^i_e=0.081$ atom$^{-1}$, respectively.

\begin{figure} 
\includegraphics[width=90mm]{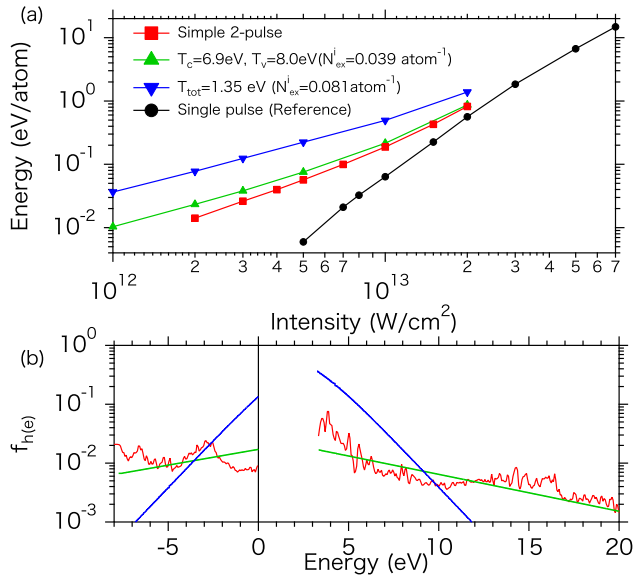} 
\caption{\label{fig5} 
(a) Electron distribution dependence of $E_{ex}$  for the second pulse. $E^{2nd}_{ex}$ produced by the first pulse is represented by black circles for reference.
(b) Filling rate of the prepared electron distribution for the second pulse.} 
\end{figure}

Figure~\ref{fig5} (a) shows the energy absorbed from the second pulse, $E^{2nd}_{ex}$.
For reference, $E_{ex}$ absorbed from the single pulse is plotted using black circles.
The parameters of the second laser pulse are same as those of the first pulse.
In the first assumption (red squares), the time delay between the two pulses was set at 19 fs.

In all cases, $E^{2nd}_{ex}$ is higher than the energy from the single pulse calculations, 
and shows a linear dependence on laser intensity.
This result indicates that  photoabsorption by the plasma  is important.
In particular, the efficiency is very high in the case where  the temperature is fixed at $T_{tot}=1.35$ eV (part of the third assumption).
In this case, the excited electron density is twice that of the other cases, (0.081 atom$^{-1}$), 
which may be the reason why this case is the most efficient.

While $N^i_e$ is the same in the other two cases, we can also see significant differences between them. 
The simple two-pulse simulation (red squares) shows the worst results, 
and the two-temperature assumption (green triangles) shows better results. 
The differences among these three cases involve the distributions of the electrons and holes when the second pulse irradiates the surface.
The filling rate, $f_{h(e)}$, before the second pulse begins is shown in  Fig.~\ref{fig5} (b).
A higher electron (hole) concentration at the  bottom (top) of the conduction (valence) band produces a higher $E^{2nd}_{ex}$. 
In particular, the simple double pulse case shows a non-monotonic distribution.
In other words, the electron excitation by the second pulse is dependent on the features of the prepared plasma.

\begin{figure} 
\includegraphics[width=90mm]{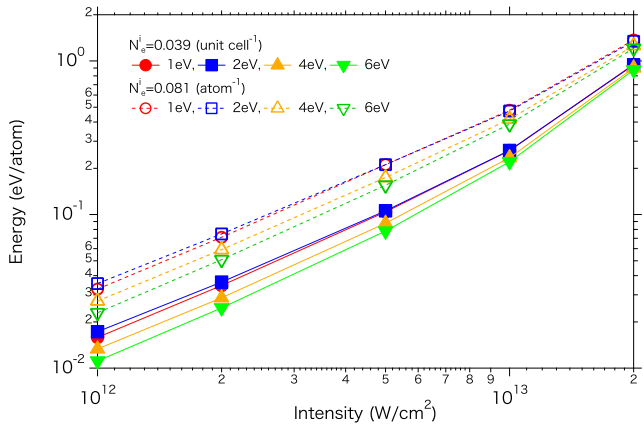} 
\caption{\label{fig6} 
Initial electron temperature dependence of  $E^{2nd}_{ex}$  on the second pulse at fixed $N^i_e=0.039$ (atom$^{-1}$) (solid lines), and $0.081$ (atom$^{-1}$) (dashed lines). } 
\end{figure}

To describe the electron temperature dependence of the electron excitation by the second pulse,
we calculated the $E^{2nd}_{ex}$ using different initial numbers of excited electrons ($N^i_e$) and the temperatures.
We define the electron temperature as $T_c=T_v\equiv T_{cv}$, to simplify the assumption.
Figure~\ref{fig6} shows the dependence of $E^{2nd}_{ex}$ on  $T_{cv}$ and $N^i_e$.
The efficiency increases as $T_{cv}$ decreases in both cases of $N^i_e$, and is highest at  $T_{cv}=2$ eV.

\begin{figure} 
\includegraphics[width=90mm]{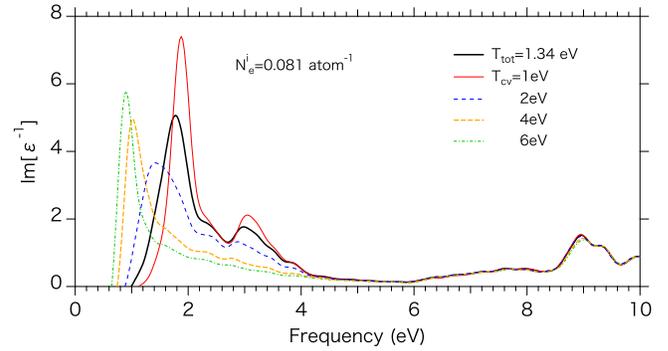} 
\caption{\label{fig7} 
$T_{cv}$ dependence of Im[$\varepsilon^{-1}$].
The number of electrons in the conduction band of the initial state is fixed at 0.081 atom$^{-1}$.
The one-temperature model at $T_{tot}=1.34$ eV is represented by the thick black line.} 
\end{figure}

One  possible reason why $E^{2nd}_{ex}$ is dependent on $N^i_{ex}$ and $T_{cv}$ is its dependence on the formed plasma. 
The most  convenient way to show the plasma contribution to
the response is to plot the imaginary part of the inverse
dielectric function Im[$\varepsilon^{-1}$]
Figure~\ref{fig7} shows the Im[$\varepsilon^{-1}$] 
for various temperature $T_{cv}$ at a fixed $N^i_{ex}=0.081$ atom$^{-1}$. 
The plasmon peak shifts from 0.8 to 2 eV as the $T_{cv}$ decreases.
According to the simple Drude model, the plasma frequency is dependent on  
$1/\sqrt{m^*}$ where $m^*$ is the effective electron mass.  
Therefore, our results indicate that the effective mass increases as the $T_{cv}$ increases, 
which is consistent with the prior work of Sato {\it et al} \cite{Sato14}.
Electrons and holes on higher energy states have heavy effective masses and this causes a  lower plasma frequency.

From first-principles  simulations, we conclude that the efficiency of the laser excitation using 
double pulses is dependent on the features of the plasma formed by the first pulse.
A higher excited electron density at a lower temperature is preferred 
because the lower temperature produces a higher plasma frequency.
While this condition appears to be conflicting in wide-gap semiconductors and insulators, 
it may be possible to produce it using laser pulse tuned to the optical band gap 
or phonon-assisted excitation of the band edge by the first pulse.

\section*{Acknowledgement}
This work was supported by  JSPS KAKENHI (Grants Nos. 15H03674 and 17K05089) and CREST (JST Grant No. JPMJCR16N5). 
Numerical calculations were performed on the SGI ICE X supercomputer at 
the Japan Atomic Energy Agency (JAEA).

\end{document}